\documentclass[a4paper]{jpconf}
\usepackage{graphicx}
\usepackage{xcolor}
\usepackage{amsmath,mathtools}
\usepackage{framed}
\colorlet{shadecolor}{lightgray}
\usepackage{hyperref}
\usepackage{parskip}

\usepackage{tensor}
\usepackage{amssymb}
\usepackage{graphicx}
\usepackage{physics}
\usepackage{lipsum}  
\usepackage{bigints}
\usepackage{verbatim}
\usepackage{amsmath, amssymb, amsthm, graphicx, epsfig, fancyhdr,epsfig}
\usepackage{float}
\usepackage{comment} 
\newcommand{\be}{\begin{equation}}
\newcommand{\ee}{\end{equation}}
\newcommand{\bea}{\begin{eqnarray}}
\newcommand{\eea}{\end{eqnarray}}

\newcommand{\non}{\nonumber}

\DeclareMathOperator{\arcsinh}{arcsinh}

\DeclareMathOperator{\arctanh}{arctanh}

\begin{document}
\title{Scale-invariant inflation}

\author{M Rinaldi$^{1, 2}$, C Cecchini$^{1, 2}$, A Ghoshal$^{3}$ and D Mukherjee$^{4}$}

\address{$^1$ {Dipartimento di Fisica, Università di Trento,
Via Sommarive 14, I-38123 Povo (TN), Italy}}
\address{$^2$ TIFPA-INFN,
Via Sommarive 14, I-38123 Povo (TN), Italy}
\address{$^3$ {Institute of Theoretical Physics, Faculty of Physics, University of Warsaw, ul. Pasteura 5, 02-093 Warsaw, Poland}}
\address{$^4$ {Department of Physics, Indian Institute of Technology Kanpur
Kalyanpur, Kanpur 208016, India}}

\begin{abstract}
We examine a scalar-tensor model of gravity that is globally scale-invariant. When adapted to a spatially flat Robertson-Walker  metric, the equations of motion describe a dynamical system that flows from an unstable de Sitter space to a stable one. We show that during this transition inflation can occur. Moreover, at the final fixed point, a mass scale naturally emerges that can be identified with the Planck mass. We compute the inflationary spectral indices and the tensor perturbation and we compare them with observations. We also study the possibility that primordial magnetic fields are generated during inflation.
\end{abstract}

\section{Introduction}
\label{intro}

The extension of the Einstein-Hilbert action to include quadratic terms in the Riemann tensor and its contractions is usually motivated by the renormalization of one-loop corrections in the context of the  semi-classical approach to quantum gravity \cite{Birrell:1982ix}. The theory\footnote{In this paper, we set $G = (8\pi M^2_{P})^{-1} = \hbar = c = 1$ and we adopt a spatially flat Robertson-Walker metric, $ds^2 = -dt^2 + a(t)^2\delta_{ij}dx^idx^j$. In conformal time, $ds^2 = -a^2(\eta)[\eta_{\mu\nu}dx^{\mu}dx^{\nu}]$. }
\bea
S=\int d^4x\sqrt{g}\left({M^2\over 2}\left(R-2\Lambda\right)+\alpha R^2+\beta R_{\mu\nu}R^{\mu\nu}\right)\,,
\eea 
has been proven to be renormalizable in 1977 \cite{Stelle:1977ry}. Once expanded on flat space, the theory shows one massless spin-2 excitation (the usual graviton), one ghost massive spin-2 excitation, and a massive scalar field. When $\beta=0$, however, ghosts are absent. When curvature is large, the quadratic terms dominate and the theory acquires a global scale-invariance. This means that the action is invariant under the dilatation $\bar g_{\mu\nu}(x)=g_{\mu\nu}(\ell x)$ for some positive and constant $\ell$. Scale-invariance is then broken at low curvature by the linear term in $R$ or, at higher energy, by quantum logarithmic corrections.

The simplest scale-invariant model of gravity is given by 
\bea\label{purequad}
S_J={\alpha\over 36}\int d^4x\sqrt{g}R^2\,.
\eea 
As such, one might see this theory as a special case of $f(R)$ gravity written in the Jordan frame \cite{DeFelice:2010aj,Sotiriou:2008rp}. By the simple conformal transformation $\tilde g_{\mu\nu}=\Omega^2 g_{\mu\nu}$, where $\Omega=\sqrt{\alpha R}/(3M)$, the action can be written in the standard Einstein frame, as a scalar-tensor theory of gravity, that is
\bea
S_E=\int d^4x\sqrt{\tilde g}\left[{M^2\over 2}(\tilde R-2\Lambda)-\frac12\tilde g^{\mu\nu}\tilde\partial_{\mu}\psi\tilde\partial_{\nu}\psi\right]\,,
\eea
where $\psi=\sqrt{6}\,M\ln\Omega$ is the scalar field that encodes the physical degree of freedom associated with $R^2$,  $\Lambda=9M^2/(4\alpha)$ and $M$ is a redundant mass parameter that does not break global scale invariance. This means that the variation of $S_E$ with respect to $M$ yields an equation that identically vanishes on-shell. The field equations in the Jordan frame and in the Einstein frame read, respectively
\bea 
RR_{\mu\nu}-\frac14 R^2 g_{\mu\nu}+(g_{\mu\nu}\Box -\nabla_{\mu}\nabla_{\nu})R=0\,,\\\nonumber
\tilde G_{\mu\nu}+\Lambda \tilde g_{\mu\nu}-M^2\left(\tilde\partial_{\mu}\psi \tilde\partial_{\nu}\psi-\tilde g_{\mu\nu}\tilde\partial_{\alpha}\psi\tilde\partial^{\alpha}\psi\right)=0\,,
\eea 
and it is evident that, while the first admits the Minkowski metric as a solution, this is not true for the second since $\Lambda$ cannot vanish. This implies that all Ricci flat solutions in the Jordan frame cannot be mapped into the Einstein frame. In these terms, the two frames are not equivalent \cite{Rinaldi:2018qpu}.

The theory \eqref{purequad} has been studied in the context of static and rotating black holes thermodynamics in \cite{Cognola:2015uva,Cognola:2015wqa}. In the framework of cosmology, it can be shown that, with a spatially flat Robertson-Walker metric and in the absence of matter, the most general solution interpolates between a radiation-dominated Universe (that is the scale factor $a(t)\propto \sqrt{t}$) and a de Sitter Universe with cosmological constant $\Lambda$, which is arbitrary as $M$ and $\alpha$. Although not very interesting for cosmology, this model reveals a possible mechanism of classical and dynamical breaking of the scale-invariance. In fact, the system evolves from an unstable configuration, where scale invariance is apparent (radiation domination), to a de Sitter space characterized by a scale $\Lambda$, which, although arbitrary, formally breaks global scale-invariance.

This mechanism becomes even more evident if one adds a non-minimally coupled scalar field with quartic potential so that scale-invariance of the action is preserved. The main focus of this work is to study this scenario, namely a scale-invariant tensor-scalar theory of gravity which was introduced in \cite{Rinaldi:2015uvu} and refined in \cite{Vicentini:2019etr,Tambalo:2016eqr}.  The principal interest in this configuration is that it shows, like in the simplest $R^2$ model, a natural transition from an unstable fixed point to a stable one but, during the process, a quasi-exponential expansion of the Universe occurs making it a viable model for inflation.

In Sec.\ \ref{secmodel} we review the model and its general dynamical properties. In Sec.\ \ref{secinflation} we explore the inflationary mechanism induced by this model and we show that it is very competitive with other models, in particular the Starobinsky one, in terms of spectral indices. In Sec. \ref{secmagnetogenesis} we also study the possible generation of primordial magnetic fields. We then conclude with some remarks and future work.

\section{Scale-invariant scalar-tensor theory}
\label{secmodel}

Following \cite{Rinaldi:2015uvu}, we consider the scale-invariant action
\begin{equation}
\label{eq:fullquadraticaction}
    S_J=\int d^4x \sqrt{-g}\left[\frac{\alpha}{36}R^2+\frac{\xi}{6}\phi^2R-\frac{1}{2}\partial_{\mu}\phi\partial^{\mu}\phi-\frac{\lambda}{4}\phi^4 \right]\ .
\end{equation}
where $\alpha,\xi,\lambda$ are dimensionless parameters that we wish to constrain by using inflationary data.
To do so, it is simpler to study the theory in the Einstein frame. Thus we perform a field redefinition and a conformal transformation that turn the action above into its Einstein frame form, namely \cite{Rinaldi:2015uvu}
\begin{equation}
\label{eq:einsteinframe-action}
    S_E= \int d^4x\ \sqrt{-{g}}\left[\frac{M^2}{2}{R}-\frac{1}{2}{g}^{\mu \nu}\partial_{\mu}{\omega}\partial_{\nu}{\omega}-\frac{1}{2}e^{-\sqrt{\frac{2}{3}}\frac{{\omega}}{M}}{g}^{\mu \nu}\partial_{\mu}\phi \partial_{\nu}\phi-V({\omega},\phi)\right]\,,
\end{equation}
where $\phi$ is the original field and the new scalar field $ \omega $ encodes the degree of freedom contained in the $R^2$ term of $S_J$.  The potential
\begin{equation}
\label{eq:potentialV}
    V({\omega},\phi)=\left(\frac{\lambda}{4}+\frac{\xi^2}{4\alpha} \right)e^{-2\sqrt{\frac{2}{3}}\frac{{\omega}}{M}}\phi^4-\frac{3M^2\xi}{2\alpha}e^{-\sqrt{\frac{2}{3}}\frac{{\omega}}{M}}\phi^2+\frac{9M^4}{4\alpha}  \ ,
\end{equation}
involves the two scalars $\phi$ and $\omega$, both minimally coupled to gravity.

Despite the potential apparently depending on two fields, a suitable scalar field space redefinition allows us to show that it really depends on one scalar degree of freedom only \cite{Tambalo:2016eqr}. This feature is exclusive to this model due to the global scale-invariance. In fact, let us define
\bea\label{eq:zeta}
    & \zeta &=\sqrt{6}M\arcsinh\left(\dfrac{\phi}{\sqrt{6}M}\,e^{-{\omega\over \sqrt{6}M}}\right) ,\\
    \label{eq:rho}& \rho &= \dfrac{M}{2}\ln \left(\dfrac{\phi^2}{2M^2} + 3e^{2\omega\over \sqrt{6}M}\right).
\eea
Then, the Einstein-frame action  becomes
\be
\label{eq:LEsinglefield}
    S_E =\int d^4x \sqrt{g}\left[ \dfrac{M^2}{2}R - \dfrac{1}{2}\left(\partial\zeta\right)^2 - 3\cosh^2\left(\dfrac{\zeta}{\sqrt{6}M}\right)\left(\partial \rho\right)^2 - U(\zeta)\right], 
\ee
where $\Omega=\alpha \lambda+\xi^2$. The potential now depends on $\zeta$ only and reads
\be
\label{eq:potential}
    U(\zeta) =  \dfrac{9\Omega M^4}{\alpha}\sinh^4\left(\dfrac{\zeta}{\sqrt{6}M}\right) -\dfrac{9\xi M^4}{\alpha}\sinh^2\left(\dfrac{\zeta}{\sqrt{6}M}\right) + \dfrac{9M^4}{4\alpha}. 
\ee
$U(\zeta)$ has two minima at 
\be
\cosh \left(\dfrac{\zeta_{\rm min}}{\sqrt{6}M}\right) = \sqrt{\dfrac{2\Omega + \xi}{2\Omega}}, 
\ee
and a local maximum at $\zeta_{\rm max} = 0$. At these points, the potential takes the following values
\be
U_{\rm max} = \dfrac{9M^4}{4\alpha}, \quad U_{\rm min} = \dfrac{9\lambda M^4}{4\Omega}. 
\ee
Note that, at the minima, the potential does not vanish. This has important consequences on the inflationary features as we show below.
By imposing the spatially flat Robertson-Walker metric $ds^2=-dt^2+a(t)^2\delta_{ij}dx^idx^j$ we obtain the Friedmann equations
\begin{align}
    \label{eq:eom1}3M^2H^2 &= \dfrac{\dot{\zeta}^2}{2} + 3\cosh^2\left(\dfrac{\zeta}{\sqrt{6}M}\right)\dot{\rho}^2 + U, \\
    \label{eq:eom2}2M^2\dot{H} &= -\dot{\zeta}^2 - 6\cosh^2\left(\dfrac{\zeta}{\sqrt{6}M}\right)\dot{\rho}^2, 
\end{align}
while the Klein-Gordon equations read
\begin{align}
    \label{eq:eom3}&\ddot{\zeta} + 3H \dot{\zeta} - \dfrac{\sqrt{6}}{2M}\sinh\left(\dfrac{2\zeta}{\sqrt{6}M}\right)\dot{\rho}^2 + \dfrac{dU(\zeta)}{d\zeta} = 0, \\
    \label{eq:eom4}&\ddot{\rho} + 3H \dot{\rho} + \dfrac{2}{\sqrt{6}M}\tanh\left(\dfrac{\zeta}{\sqrt{6}M}\right)\dot{\zeta}\dot{\rho} = 0.
\end{align}
As it is apparent, $\rho$ = constant is one solution of the system and it can be shown that it is also an attractor. In other words, the dynamics is driven by the metric and the field $\zeta$ only. Again, this feature descends from the scale-invariance of the theory.

To prove that $\rho$ = constant is an attractor solution we first convert the time derivative into derivatives with respect to $N=\ln a$. Then, equation \eqref{eq:eom2} and the two equations \eqref{eq:eom3},\eqref{eq:eom4} become, respectively
\bea\label{newdynsys}
H'&=&-H\left({A^2\over 2M^2}+{3C(\zeta)^2\sigma^2\over M^2}\right)\,,\\\non\\\non 
A'&=&-\left(3-{A^2\over 2M^2}-{3C(\zeta)^2\sigma^2\over M^2}\right)A+{\sqrt{6}\over 2M }S(\zeta)\sigma^2-{1\over H^2}{dU\over d\zeta}\,,\\\non\\\non 
\sigma'&=&-\left(3-{A^2\over 2M^2}-{3C(\zeta)^2\sigma^2\over M^2}+{2T(\zeta)A\over\sqrt{6}M}\right)\sigma\,,
\eea
where we have defined the two new variables $A,\sigma$ through 
\bea\label{newdef}
\rho'&=&\sigma\,,\\\non\\\non 
\zeta'&=&A\,.
\eea
Here
  $C(\zeta)=\cosh\left[\zeta/( \sqrt{6}M)\right]$, $T(\zeta)=\tanh\left[\zeta/( \sqrt{6}M)\right]$, $S(\zeta)=\sinh\left[2\zeta/( \sqrt{6}M)\right]$, 
  and the prime is the derivative with respect to $N$. The system \eqref{newdynsys} and \eqref{newdef} is now composed of five non-linear first-order equations whose fixed points are determined by solving the algebraic system $(H',A',\sigma',\rho',\zeta')=(0,0,0,0,0)$, which yields the two solutions
\bea\label{stab}
A=0\,,\quad \sigma=0\,,\quad \zeta=\sqrt{6}M\,{\rm arccosh}\left(\sqrt{2\Omega+\xi\over 2\Omega}\right)\,,\quad H_{\rm st}={M\over 2}\sqrt{3(\Omega-\xi^2)\over \alpha\Omega}
\eea
and
\bea\label{unst} 
A=0\,,\quad \sigma=0\,,\quad \zeta=0\,,\quad H_{\rm unst}={\sqrt{3}M\over 2\sqrt{\alpha}}\,. 
\eea 
Here we have also used the first Friedmann equation \eqref{eq:eom1} to compute explicitly the values of $H$ and $\zeta$.
By linearizing the right-hand-side of the system around these solutions, we see that the equation for $\sigma$ is just $\sigma'=-3\sigma$ thus $\sigma=$ 0 (that is $\rho=$ const) is a stable attractor in both cases.
By solving the remaining linearized equations it is easy to show that the fixed point \eqref{stab} is a stable attractor while the fixed point \eqref{unst} turns out to be a saddle point. The analysis of the system confirms that $\rho$ is nothing but an arbitrary constant during the evolution of the system that connects two de Sitter spaces. The Hubble parameter decreases from $H_{\rm unst}$ to $H_{\rm st}$ and, at the same time, the field $\zeta$ increases from a vanishing value at the unstable fixed point, where the potential has a local maximum, to reach a finite value where the potential has a non-vanishing minimum.

In terms of the original Jordan frame formulation, the scalar field $\phi$ vanishes at the unstable point while it reaches a constant value $\phi_{\rm st}$ at the fixed point. In \cite{Rinaldi:2015uvu} it was shown that, by making the choice $\alpha=\xi^2/\lambda$, the action \eqref{secmodel} reduces to the Einstein-Hilbert one at the stable fixed point because the quadratic term in $R$ asymptotically cancels the quartic term in $\phi$, provided one makes the identification $M_{\rm pl}=\sqrt{\xi/3}\,\phi_{\rm st}$, where $M_{\rm pl}$ is the Planck mass (since $\phi$ becomes constant also the kinetic term in $\phi$ vanishes). In this sense, the model breaks scale-invariance dynamically since at the end of the evolution the scalar fields stabilise around a non-vanishing value that can be related to the fundamental Planck mass. When  $\alpha$ is a free parameter, as in the case at hand here, the action \eqref{secmodel} still reduces to the Einstein-Hilbert one with negligible corrections at low energy. This is because, as we will shortly see, inflationary observables constrain the parameter $\alpha,\xi,\lambda$ into a small range very close to (although not exactly equal to) $\alpha=\xi^2/\lambda$.

\section{Inflation}
\label{secinflation}
In the previous section, we showed that the scale-invariant action \eqref{secmodel} can be written in the Einstein frame in such a way that the potential depends on one field only, see eq.\ \eqref{eq:LEsinglefield}. We also showed that, with a spatially flat Robertson-Walker metric, the resulting equations of motion describe a transition between two de Sitter spaces, one unstable and one stable. We now show that during the transition inflation can occur provided the parameters satisfy certain constraints.

 The first two slow-roll parameters obtained from \eqref{eq:potential} read
\bea\label{e1}
\epsilon&=&{M^{2}\over 2}\left({1\over U}{dU\over d\zeta}\right)^{2}={16y(y+1)(2\Omega y-\xi)^2\over 3(4\Omega y^2-4\xi y+1)^2}\,,\\\non\\\label{e2}
\eta&=&{M^{2}\over U}{d^{2}U\over d\zeta^{2}}={4\left[8\Omega y^2+2y(3\Omega -\xi)-\xi\right]\over 3(4\Omega y^2-4\xi y+1)}\,,
\eea
where we set
\bea
y=\sinh^2\left(\zeta\over\sqrt{6}M\right)\,.
\eea 
The complexity of their expressions makes it inevitable to perform some approximation. In particular, it is convenient to distinguish between small ($\zeta\ll M$) and large ($\zeta\geq M$) field regimes.

In the first case, the potential can be approximated with
\bea
\label{eq:smallfieldV}
U \approx {9M^4\over 4\alpha}\left[1-{2\xi\over 3}{\zeta^2\over M^2}-\frac{1}{9}\left({\xi\over 3}-\Omega\right){\zeta^4\over M^4}+{\cal O}\left(\zeta^6\over M^6\right)\right]
\eea
which has the form of hilltop inflation. The analysis of the Cosmic Microwave Background (CMB) performed by the Planck Mission has shown that the quartic part of the potential must dominate with a negative coefficient \cite{Planck:2018jri}. Thus $\xi$ should be very small but then the coefficient of the quartic term would be positive thus contrasting observations. Therefore, we exclude the small field limit as a viable inflationary model.

\begin{figure}[h]
	\centering 
	\includegraphics[scale=0.45]{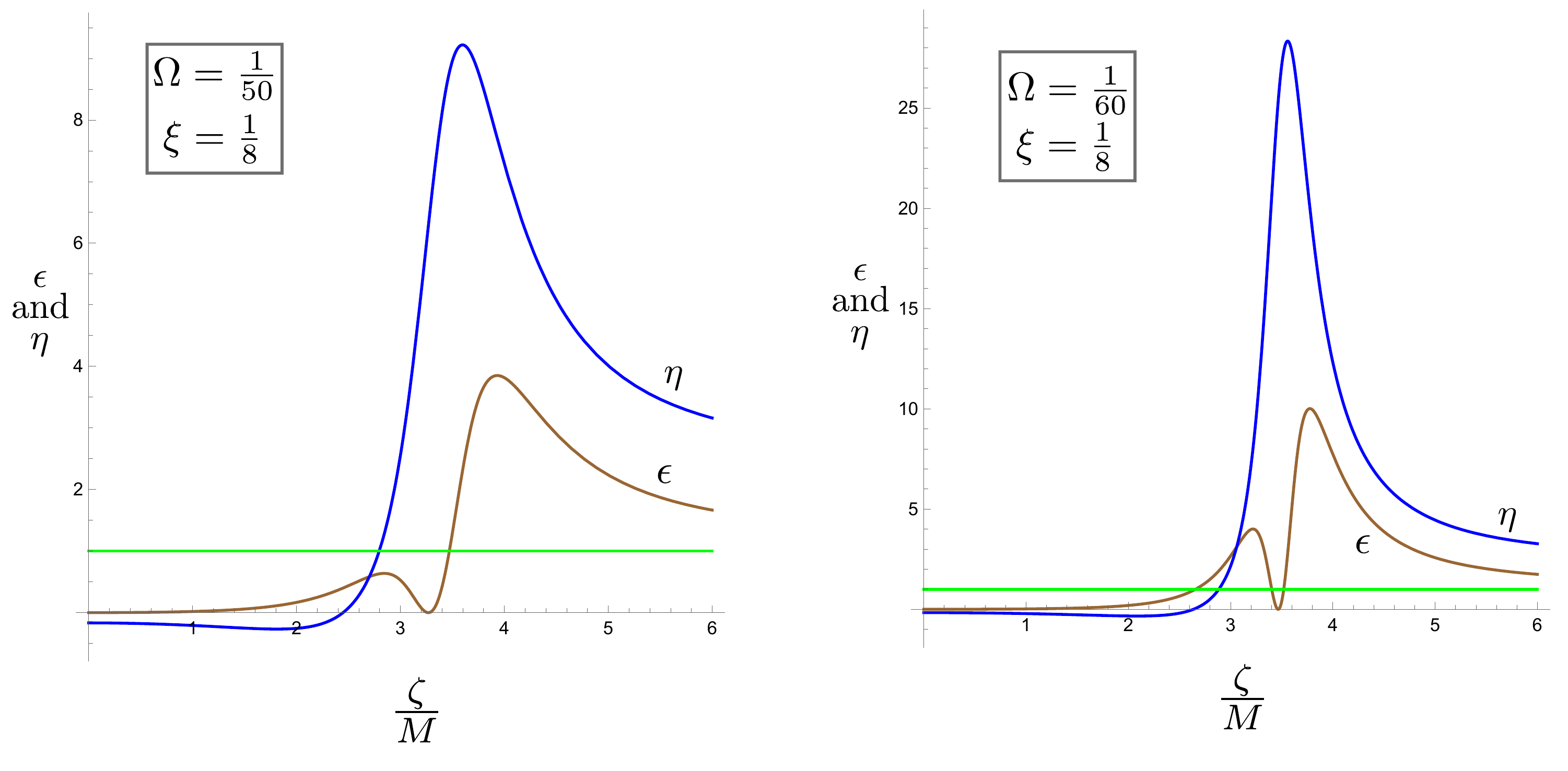} 
	\caption{Plot of the spectral indices $\eta$ and $\epsilon$ with respect to the dimensionless quantity $\zeta/M$ for $\Omega=1/50$ and $\xi=1/8$ on the left and $\Omega=1/60$ and $\xi=1/8$ on the right.}
	\label{etaVeps}
\end{figure}

We now turn to the large field limit. In this case, there is a further restriction on the parameter space. First of all, the analysis of the expressions for the slow-roll parameters shows that $\Omega<1/64$ to avoid divergences \cite{Ghoshal:2022qxk}. Moreover, for a certain combination of $\Omega$ and $\xi$ the slow roll parameter $\epsilon$ vanishes before reaching the benchmark for the end of inflation, that is $\epsilon\simeq 1$ (see plot \ref{etaVeps}). 
This is problematic because the duration of inflation is given by the standard formula
\bea\label{deltan}
\Delta N=-{1\over M}\int_{\zeta_*}^{\zeta_{\rm end}}{d\zeta\over \sqrt{2\epsilon}}\,,
\eea
where $\zeta_*$ and $\zeta_{\rm end}$ mark the beginning and the end of inflation. Clearly, if $\epsilon$ vanishes in between the integral diverges and inflation lasts forever. The vanishing of the slow-roll parameter is due to the fact the potential looks like a Mexican hat one, with the crucial difference though that at the minima it does not vanish since $U_{\rm min}=9M^4\lambda/(4\Omega)$.

By taking into account the constraints above, it turns out that inflation can occur during the journey of the field $\zeta$ from the local maximum (unstable fixed point), where it is vanishing, to the local minimum of the potential where it reaches a finite value (stable fixed point).

In the large field limit the potential can be approximated by
\bea\label{utilde}
U={9M^4\over 4\alpha}\left[1-\xi\,\exp\left({\sqrt{6}\zeta\over 3M}\right)+{\Omega\over 4}\,\exp\left({2\sqrt{6}\zeta\over 3M}\right)\right]\,.
\eea
The first slow-roll parameter reads
\bea\label{epsx}
\epsilon(x)={4x^{2}(\Omega x -2\xi)^{2}\over 3(\Omega x^{2}-4\xi x +4)^{2}}\,,
\eea
where 
\bea
x=\exp\left(\sqrt{\frac23}{\zeta\over M}\right)\gg 1\,.
\eea
The equation $\epsilon=1$ has four solutions. If we want to avoid that $\epsilon$ vanishes before reaching unity (that is the situation in the left panel of fig. \ref{etaVeps}), we need them to be all real, which requires \cite{Ghoshal:2022qxk}
\bea\label{Omegaconstr}
\Omega<{2\sqrt{3}\over 3}\xi^{2}\simeq 1.1547\, \xi^{2}\,.
\eea
This bound must be combined with $\Omega>\xi^{2}$, which guarantees that the potential is positive definite. The result is that  $0<\alpha\lambda<0.1547\,\xi^{2}$. In addition, the constraint $\Omega<1/64$, which prevents the slow-roll parameters to diverge, finally implies that $\xi\lesssim 0.11$, which is much smaller than the values predicted by Higgs inflation at the classical level.

The spectral indices read
\bea
&&n_{s}={48-5x^4\Omega^2-8\Omega\xi x^3+(16\xi^2+88\Omega)x^2-160\xi x\over 3(\Omega x^2-4 x\xi+4)^2}\,,\\\non
&&r={64x^2(\Omega x-2\xi)^{2}\over 3(\Omega x^2-4x\xi+4)^2}\,.
\eea
If we invert  $r$ to find $x=x(r)$, we choose the solution corresponding to the first positive zero of $\epsilon=1$ and expand for small $r$ we find
\bea
x_{r}\simeq {\sqrt{3r}\over 4\xi}\,.
\eea
By substituting $x$ with $x_{r}$ in $n_{s}$ and expanding again for small $r$ we find
\bea\label{nsofr}
n_{s}\simeq 1-\sqrt{r\over 3}\,,
\eea
which is the same relation found in the Starobinsky model. As a last step, one needs to compute the field value at the beginning of inflation. To do so one needs to invert the relation \eqref{deltan}. It can be shown then that the initial value of the field $x_i$ is related to the final one $x_f$ (computed by considering the smallest positive root of the equation $\epsilon=1$ with $\epsilon$ expressed by eq.\ \eqref{epsx}) and to the desired number of e-foldings $\Delta N$ by the formula
\bea
x_i=-{2\over \xi}\,{ W_{-1}^{-1}\left[ -{2\over \xi x_f}\,\exp\left(  -\frac43 \Delta N -{2\over \xi x_f}\right) \right]}\,,
\eea
where $W_{-1}$ is the lower branch of the Lambert function \cite{Ghoshal:2022qxk}.

For instance, if $\Delta N=60$, $\Omega=1.07\, \xi^{2}$ and $\xi=10^{-2}$, we find
\bea\label{numbers}
x_{i}=2.33\rightarrow \zeta_{i}=1.04\,M\,,\quad x_{f}=99.78\rightarrow \zeta_{f}=5.637\,M\,,
\eea
which implies that the approximation $\zeta\geq M$ is appropriate. In addition, we also find
\bea\label{indices}
n_s=0.9679\,,\quad r=0.003\,,\quad \xi_{v}^{2}=0.00024\,,
\eea
which are fully compatible with Planck data. 

In the following figure, we plot the $r$-vs-$n_s$ curve obtained from our model comparing it with constraints obtained for these cosmological observables from Planck, BICEP. It also remains within the range of detectability of future detectors such as Simons Observatory.
\begin{figure}[H]
	\centering 
	\includegraphics[scale=0.48]{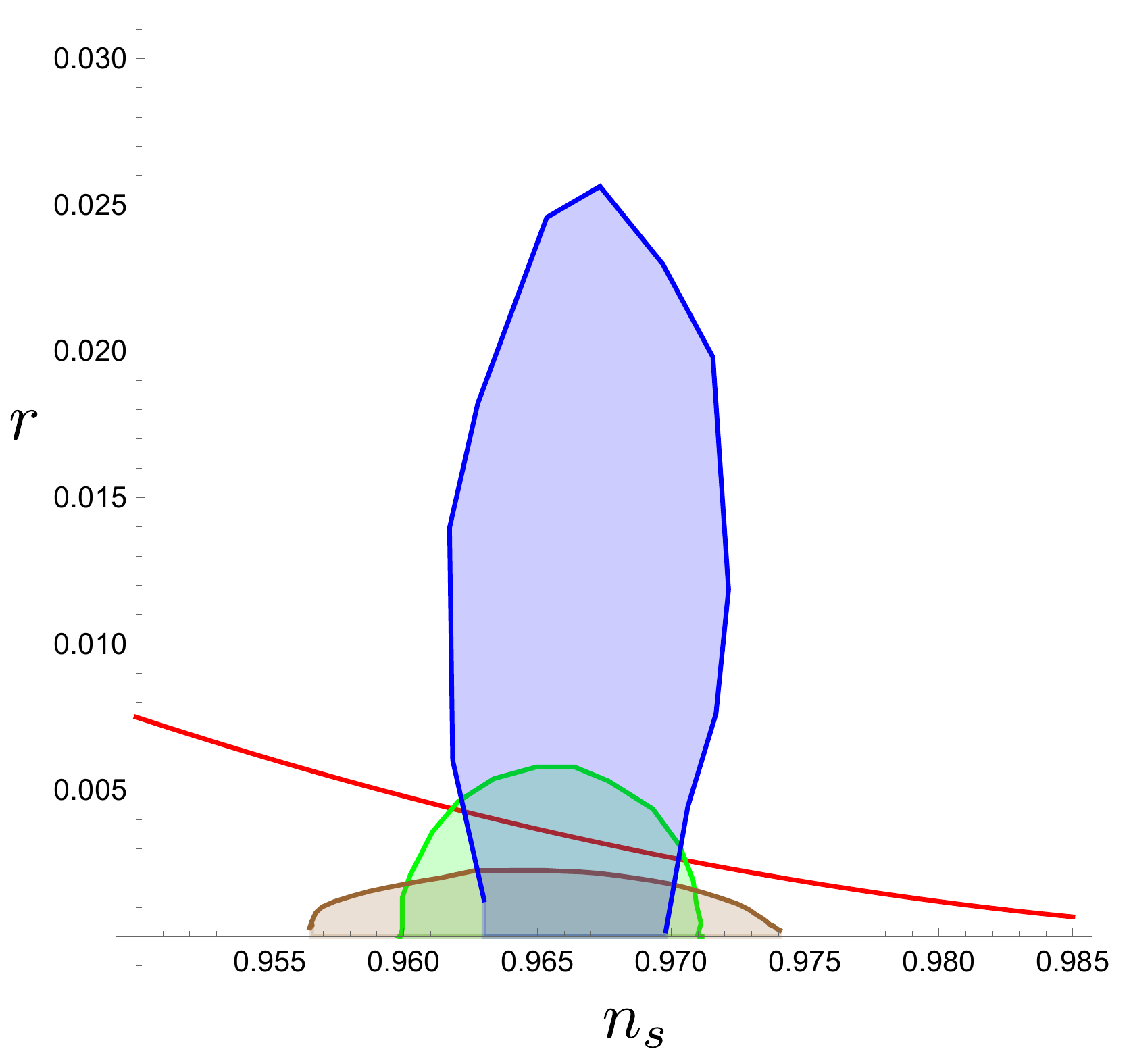} 
	\caption{\it Plot of $r$ vs $n_s$ following from \eqref{nsofr}. The blue region represents the current PLANCK + BICEP constraints \cite{Planck:2018vyg, BICEP:2021xfz, BICEPKeck:2022mhb, Campeti:2022vom}
, the green region represents the future reach of Simon's Observatory\cite{SimonsObservatory:2018koc}  while the brown region depicts the detection range of LiteBIRD \cite{LiteBIRD:2022cnt}}
	\label{nsrdata}
\end{figure}

\subsection*{\textbf{Tensor perturbations}}

The analysis of tensor perturbations and the tensor power spectrum in the model under consideration mimics the \emph{standard} case where the gravitational sector is governed by the usual Einstein-Hilbert action. Turning on tensor perturbations $h_{ij}$, we write down the perturbed metric as
\begin{equation}
\label{eq:pertmetric-gaugeexplicit}
    ds^2= a(\eta)^2\left[-d\eta^2+(\delta_{ij}+{h}_{ij})dx^idx^j \right]\ ,
\end{equation}
and restrict ourselves to the transverse-traceless gauge given by
\begin{equation}
    \partial_i h^{ij}=h=0\ .
\end{equation}
The equations of motion for the tensor perturbations are given by,
\begin{equation}
\label{eq:final-pert-eom}
    2M^2\partial_{\eta}^2 h_{ij}+4M^2 \frac{a'}{a}\partial_{\eta}h_{ij}-2M^2\partial^k \partial_k h_{ij}=0 .
\end{equation}
where primes are derivatives with respect to the conformal time $\eta$ defined as
$\eta=\int \frac{dt}{a(t)}$. The above equation is consistent with the tensor perturbations for any theory of inflation with Einstein gravity coupled to other matter fields in a \emph{minimal} way. This is somewhat expected since, although we started with higher derivative terms in the gravitational sector, we effectively mapped the system to an Einstein frame action with two scalars given by \eqref{eq:einsteinframe-action}. This trade-off essentially simplifies the tensor perturbations in the gravitational sector at least up to linear order. 
Performing a momentum space analysis following \cite{Baumann:2009ds}, and quantizing the fluctuations, we see that the tensor power spectrum takes the form
\begin{equation}\label{powerspectrum}
    \mathcal{P}_h= \frac{2H_{*}^2}{\pi^2 M^2}\ ,
\end{equation}
where the right-hand side is measured at horizon crossing. The power spectrum obtained is in fact universal for inflationary scenarios whose gravitational sector is described exclusively by the Einstein-Hilbert action.

Strictly speaking the Einstein frame action \eqref{eq:einsteinframe-action} consists of two scalars $\phi$ and ${\omega}$ which might motivate someone to perform a multi-field inflationary analysis. However, as argued earlier, a field redefinition renders the potential dependent on only one of the fields while the other direction $\rho$ can be treated as a \emph{flat} direction and does not affect inflationary dynamics. The analysis of tensor perturbations indeed shows that the Friedman equations \eqref{eq:eom1} and \eqref{eq:eom2} conspire in a way to give us the equation of motion of $h_{ij}$ which matches with that of a single field inflationary model with Einstein-Hilbert gravity thus vindicating our analysis based on single-field inflationary models.

\section{Inflationary generation of helical magnetic fields}
\label{secmagnetogenesis}
Magnetic fields are ubiquitous in the Universe. 
Galaxies and clusters of galaxies host magnetic fields with strength from few to tens of $\mu$G and coherent on scales of tens kpc, independently on their redshift. 
Further, gamma-ray observations of distant blazars have inferred the presence of magnetic fields with large coherence lengths $L\gtrsim$1 Mpc in the cosmic voids of the intergalactic medium (IGM). Together with upper bounds from CMB observations and ultra-high-energy cosmic rays, they constrain the amplitude of IGM magnetic fields coherent on Mpc scales or larger as $10^{-16}\lesssim B_0\lesssim 10^{-9}$ G. The lower bound is larger by a factor $(L/1\, \mathrm{Mpc})^{-1/2}$ for $L<$1 Mpc. 
Moreover, non-vanishing parity-odd correlators of gamma-ray arrival directions observed by Fermi-LAT suggest that intergalactic magnetic fields are helical. 
For updated reviews on observational constraints, see, for instance, \cite{paoletti_2019, vachaspati_2021}.

The origin of such fields (magnetogenesis), however, is still controversial.
Among all the proposed scenarios, the most promising one is inflation, which could naturally lead to large coherence lengths and explain all at once galactic, cluster, and IGM magnetic fields. 
For a review, see \cite{subramanian_2016}. 
As for primordial scalar and tensor perturbations, quantum vacuum fluctuations of the gauge field could have been amplified and converted into observable electromagnetic (EM) fields. 
Nevertheless, the conformal invariance of standard electromagnetism prevents the magnetic field from amplification in a spatially flat Robertson-Walker background, washing out its amplitude as $B\sim a^{-2}$. 
Inflationary magnetogenesis requires a breaking of conformal symmetry in the EM sector. 

In this section, we give an overview of the model of magnetogenesis studied in \cite{cecchini_2023}, focusing on its application to scale-invariant gravity. Essentially, the same action of the hybrid axion-Ratra model discussed in \cite{caprini_2014} is considered, 
\begin{equation}
\label{eq:axionRatra}   
     S = -\dfrac{1}{16\pi}\int d^4x\sqrt{-g}\,  I^2[\zeta(a)]\left(F_{\mu\nu}F^{\mu\nu} - \gamma F_{\mu\nu}\tilde{F}^{\mu\nu}\right) + S_E.
\end{equation}
where $F_{\mu\nu} = \partial_{\mu}A_{\nu} - \partial_{\nu}A_{\mu}$ is the strength of a $U(1)$ gauge field and $\tilde{F}^{\mu\nu} = (1/2)\epsilon^{\mu\nu\alpha\beta}F_{\alpha\beta}$ is its dual with the totally antisymmetric tensor $\epsilon^{\mu\nu\alpha\beta}$ defined as $\epsilon^{\mu\nu\alpha\beta} = 1/\sqrt{-g}\,\eta^{\mu\nu\alpha\beta}$. $\eta^{\mu\nu\alpha\beta}$ is the Levi-Civita symbol with values $\pm 1$. $\gamma$ is a positive dimensionless constant. 
The dual field strength $\tilde{F}^{\mu\nu}$ has been added to generate fully helical magnetic fields. 
The gauge field is assumed to be a test field with no influence on the scalar field evolution or spacetime geometry.

The function $I$ is in general required to depend on time, physically realized by coupling a scalar field with non-trivial dynamics to the EM sector, which breaks conformal invariance. It is known that $I$ as written in \eqref{eq:axionRatra} plays the role of an inverse coupling constant \cite{demozzi_2009}. Therefore, the coupling function should satisfy $I \gtrsim 1$ throughout inflation and approach unity in the end, thereby avoiding strong coupling (i.e., keeping the theory in its perturbative regime) and restoring conformal invariance once inflation is over. 
In \cite{cecchini_2023}, we have shown that when $I$ is modelled as a sawtooth coupling, a viable magnetogenesis is realized: the problems that are known to plague models with a monotonic function $I$ -- back-reaction or strong coupling --  are avoided, and current magnetic field's properties in agreement with observational bounds are obtained.
By introducing a sharp transition at $a = a_*$, we parametrize the coupling function as
\begin{equation}
\label{eq:sawtooth}   
I = 
\begin{cases} 
       \mathcal{C}\left(\dfrac{a}{a_*}\right)^{\nu_1}& a_i<a<a_*\\
       \mathcal{C}\left(\dfrac{a}{a_*}\right)^{-\nu_2}& a_*<a<a_f 
   \end{cases}
\end{equation}
where $\nu_i$ are positive exponents, while $a_i$ and $a_f$ denote the scale factor at the beginning and end of inflation, respectively.
The constant $\mathcal{C}$ and the scale factor at transition $a_*$ are found by imposing $I_i = I_f = 1$; in this way, EM conformal invariance is only broken during inflation.
We have further provided the model with a robust physical interpretation by linking the time dependence of $I$ to the evolution of the inflaton field of the scale-invariant model \eqref{eq:fullquadraticaction}, $\zeta$. For this reason, $S_E$ is precisely the Einstein frame action \eqref{eq:LEsinglefield}. 

In the slow roll approximation and further expanding for $\Omega\to \xi^2$ and $\xi\ll 1$ according to the analysis presented in Section \ref{secinflation}, the solution to the Klein-Gordon equation \eqref{eq:eom3} reads 
\begin{equation}
    \zeta(N) \propto \sqrt{6}M \arctanh\left[\exp\left(\frac{4\xi}{3}N\right)\right], 
\end{equation}
where the proportionality constant is set by the initial conditions. We conclude that, under these approximations, the following function reproduces the branches of the sawtooth coupling \eqref{eq:sawtooth} 
\begin{equation}
    I\left[\zeta(a)\right]\sim \tanh\left(\dfrac{\zeta(a)}{\sqrt{6}M}\right)^{\pm \frac{3}{4\xi}\nu_i} \sim a^{\pm \nu_i}.
\end{equation}

In the Coulomb gauge, $A_0 = \partial_iA^i = 0$, the Maxwell equations get modified as 
\begin{equation}
\label{eq:Maxwellgeneral}
    \mathcal{A}_{\pm}''(k, \eta) + \left(k^2 \pm 2\gamma k \dfrac{I'}{I}-\dfrac{I''}{I}\right)\mathcal{A}_{\pm}(k, \eta)=0, 
\end{equation}
where we have introduced the canonically normalized field $\mathcal{A}_{\pm} = I(\eta) A_{\pm}(k, \eta)$, and $A_{\pm}(k, \eta)$ are the Fourier modes of the quantized gauge field. The prime denotes a derivative with respect to conformal time.

Solutions to \eqref{eq:Maxwellgeneral} are found in the super-Hubble limit $(-k\eta)\ll 1$ in each stage of inflation by imposing the continuity of the physical vector potential $A_{\pm}(k, \eta)$ and its conformal time derivative at transition. See \cite{cecchini_2023} for a detailed derivation. Once the solutions $\mathcal{A}_{\pm}$ are known, it is possible to evaluate the power spectra in each stage. The electric and magnetic energy density per logarithmic interval in $k$-space are 

\begin{eqnarray}
    \label{eq:PE1}&\dfrac{d\rho_{E, I}^{\pm}}{d\ln k}& = \dfrac{d\rho_{B, I}^{\pm}}{d\ln k} \propto e^{\pm \pi \gamma \nu_1} H^4 (-k\eta)^{4-2\nu_1},\\
    \label{eq:PE2}&\dfrac{d\rho_{E, II}^{\pm}}{d\ln k}& \propto e^{\pm \pi \gamma \nu_1} H^4(-k\eta)^{6-2\nu_1}\left(\dfrac{\eta_*}{\eta}\right)^{2-2\nu_1+2\nu_2},\\
    \label{eq:PB2}&\dfrac{d\rho_{B, II}^{\pm}}{d\ln k}& \propto e^{\pm \pi \gamma \nu_1}H^4(-k\eta)^{8-2\nu_1}\left(\dfrac{\eta_*}{\eta}\right)^{2-2\nu_1+2\nu_2}.
\end{eqnarray}
As a crucial remark, when $\nu_1 = 2$ the spectra are scale-invariant ($k$-independent) in the first stage of inflation and very blue-tilted after transition, namely $d\rho^{\pm}_{E, II}/d\ln k \sim k^2$ and $d\rho^{\pm}_{B, II}/d\ln k \sim k^4$. This shows that having a sawtooth coupling is incompatible with a scale-invariant magnetic power spectrum throughout inflation. 

From equations \eqref{eq:PE1} and \eqref{eq:PB2} we can compute the magnetic field's observables. By considering a purely de Sitter spacetime, the energy density stored in the EM field at a given time $\eta = -1/aH$ is
\begin{equation}
\label{eq:rhoEM}
    \rho_{EM} (a) = \sum_{(+, -)} \rho^{\pm}_E(a) + \rho^{\pm}_B(a) = \sum_{(+, -)} \int_{k_i = a_i H}^{k = aH} \dfrac{dk}{k}\left(\dfrac{d\rho^{\pm}_E}{d\ln k} + \dfrac{d \rho^{\pm}_B}{d \ln k}\right), 
\end{equation}
where we have considered that at the moment $\eta_k$, when the corresponding wave crosses the Hubble scale, the scale factor is $a_k \simeq k/H$. 
We define the scale-average magnetic field amplitude, 
\begin{equation}
    B = \sqrt{\dfrac{8\pi}{I^2}\rho_B^{\pm}}, 
\end{equation}
where $\rho_B$ follows from integrating equations \eqref{eq:PE1}-\eqref{eq:PB2} (a detailed calculation is given in \cite{cecchini_2023}). Following \cite{durrer_2013}, we define the comoving characteristic scale of the magnetic field, which is sometimes referred to as ``correlation scale'', 
\begin{equation}
    L_c = \dfrac{2\pi}{\rho^{\pm}_B}\int \dfrac{dk}{k^2} \dfrac{d\rho^{\pm}_B}{d\ln k}. 
\end{equation}
In addition, we conveniently define the magnetic spectral index $n_B$ via 
\begin{equation}
    \dfrac{d\rho^{\pm}_B}{d\ln k} \propto k^{2 n_B}. 
\end{equation}
The evolution of $(B, L)$ after inflation is significantly affected by the presence of helicity. Note that the factors $\e^{\pm \pi \gamma \nu_1}$ in \eqref{eq:PE1}-\eqref{eq:PB2} display the effect of the parity breaking term $F_{\mu\nu}\tilde{F}^{\mu\nu}$ in the action \eqref{eq:axionRatra}: the positive polarization mode is exponentially amplified over the negative one, resulting in fully helical magnetic fields at the end of inflation. 
Once inflation is over, the electric field is shorted out due to the high conductivity of the Universe. On the other hand, the magnetic field, being helical, undergoes inverse cascade evolution: in the comoving frame, its intensity decreases, and its correlation scale increases while power is transferred from small to large scales. The following relations determine the current values of $(B, L)$
\begin{eqnarray}
    &B_0 \simeq 10^{-8}\mathrm{G}\left(\dfrac{L_0}{\mathrm{Mpc}}\right),\\
    &B^2_0L_0 = B^2_f L_f\left(\dfrac{a_f}{a_0}\right)^3, 
\end{eqnarray}
where subscripts $f$ and $0$ denote the end of inflation and the present time, respectively. 

To investigate the back-reaction problem properly, we must consider how the equations of motion \eqref{eq:eom1}-\eqref{eq:eom3} governing the inflationary dynamics are affected by the coupling to the EM sector. 
The first Friedmann equation \eqref{eq:eom1} gets modified as
\begin{equation}
    3H^2M^2 = \dfrac{1}{2}\dot{\zeta}^2 + 3\cosh^2\left(\dfrac{\zeta}{\sqrt{6}M}\right)\dot{\rho}^2 + U(\zeta) + \rho_{EM}, 
\end{equation}
with $\rho_{EM}$ as defined in \eqref{eq:rhoEM}. The Klein-Gordon equation for the inflaton \eqref{eq:eom3} becomes
\begin{equation}
    \ddot{\zeta} + 3H \dot{\zeta} - \dfrac{\sqrt{6}}{2M}\sinh\left(\dfrac{2\zeta}{\sqrt{6}M}\right)\dot{\rho}^2 + \dfrac{dU(\zeta)}{d\zeta} = \mathcal{S}_{EM}, 
\end{equation}
where we have defined
\begin{equation}
    \mathcal{S}_{EM} \equiv -\dfrac{1}{8\pi} I \dfrac{dI}{d\zeta}\left(F_{\mu\nu}F^{\mu\nu}-\gamma F_{\mu\nu}\tilde{F}^{\mu\nu}\right). 
\end{equation}
To avoid back-reaction, we require that the EM field does not spoil the coupled equations of motion for the inflaton and the background geometry. Equivalently, the following constraints must be satisfied at the same time throughout inflation
\begin{equation}
    \rho_{EM} \ll 3H^2M^2, 
\end{equation}
\begin{equation}
    \abs{\mathcal{S}_{EM}} \ll 3 H\dot{\zeta}.
\end{equation}
To derive the present-day observables $(B_0, L_0)$, we fix the inflationary dynamics first: the total duration of inflation $\Delta N = N_f - N_i$ and the parameters $\xi$ and $\Omega$ are chosen according to the analysis presented in Section \ref{secinflation}. We then fix $\nu_1 = 2$ to reproduce a scale-invariant magnetic/electric spectrum in the first stage. The remaining parameters are: the exponent $\nu_2$ characterizing the coupling $I$ in the second stage, the energy scale of inflation regulated by the overall height of the inflaton's potential through $\alpha$, and the coupling to helicity $\gamma$ as defined in \eqref{eq:axionRatra}. These quantities are varied to avoid back-reaction and provide $(B_0, L_0)$ in agreement with the observational bounds. 
The results are given in Table \ref{tab:results}.
Having $\nu_1 = 2$ implies that the EM spectra are scale-invariant in the first stage of inflation, i.e., in the large-wavelength sector probed by \textit{Planck}, as it happens with scalar and vector perturbations. 
The magnetic power spectrum in the second stage, on the other hand, is blue-tilted as $d\rho^{\pm}_{B, II}/d \ln k\sim k^4$. This scenario is further supported in the context of the inflationary background considered here: it is reasonable to think about a magnetic power spectrum reflecting the scale invariance of the underlying model at the beginning of inflation, which is replaced by a scale-dependent one when the inflaton approaches the end of the plateau of the potential and moves towards the minimum. 
\begin{center}
\begin{table}[h]
\centering
\caption{\label{tab:results}Present day magnetic field amplitude and correlation scale for different choices of the energy scale $\rho^{1/4}_{inf}$ and duration $\Delta N$ of inflation, exponents $\nu_i$, and coupling to helicity $\gamma$.} 
\begin{tabular}{c c c c c c c c}
\br
$\rho^{1/4}_{inf}\, \, [\mathrm{GeV}]\, \, $ & $ \Delta N $ & $ \nu_1 $ & $ \nu_2 $ & $ \gamma $ & $ L_0\, \, [\mathrm{Mpc}]  $ & $ B_0(L_0)\, \, [\mathrm{nG}]  $ & $B_0(\ell = 1\, \mathrm{Mpc})\, \, [\mathrm{nG}]$\\
\mr
      $3\times 10^{15}$ & 60 & 2 & 1.3 & 1 & 0.74 & 7.4 & 4.0 \\
      $5\times 10^{14}$ & 60 & 2 & 1.4 & 1 & 0.13 & 1.3 & $2.0\times 10^{-2}$ \\
      $3\times 10^{13}$ & 60 & 2 & 1.5 & 1 & $6.7 \times 10^{-3}$ & $6.7\times 10^{-2}$  & $3.0\times 10^{-6}$ \\
      $3\times 10^{13}$ & 70 & 2 & 1.5 & 1 & 0.19 & 1.9 & $6.7\times 10^{-2}$ \\
\br
\end{tabular}
\end{table}
\end{center}
All the listed results for $\nu_1 = 2$ are compatible with the window between gamma-ray lower bounds and CMB upper bounds and they are free from back-reaction. At variance with the case of a monotonic coupling $I$ \cite{caprini_2014}, these results are obtained without lowering dramatically $\rho_{inf}$, resulting at the same time in higher $(B_0, L_0)$. As a side note, in \cite{cecchini_2023} we have provided a rough estimate of the contribution to the tensor perturbations by the gauge field and we have verified that under the choice of parameters given in Table \ref{tab:results}, it is negligible if compared to the standard vacuum fluctuations. 

\section{Discussions and Conclusion}
In this review, we have analysed the most general scale-invariant theory with a scalar field written in the Einstein frame and also investigated the problem of magnetogenesis in the context of scale-invariant quadratic gravity.

Through a dynamical system analysis of the scale-invariant theory \eqref{eq:fullquadraticaction} in the Einstein frame \eqref{eq:einsteinframe-action}, we have found that the system evolves from an unstable de Sitter space to another stable asymptotic de Sitter space point, where an effective mass scale emerges, which can be identified with the Planck mass. The transition between the two points yields an inflationary expansion of the Universe. In the Einstein frame, the dynamics seems to be ruled by two scalar fields, however by means of a suitable field redefinition, the effective potential can be made to depend on \emph{one of the scalars only}, namely $\zeta$, thus reducing it to an effective single-field inflation model. The effective scalar field driving the inflation is approximately the same order as the new mass scale $M$ (identified with the Planck mass). In this limit, the spectral indices and their running show excellent agreement with observations. In addition, the three free parameters of the theory are significantly constrained by data but not fine-tuned (see \eqref{Omegaconstr}). 

These general results have been applied to a specific concrete model where the  scalar sector features the field $\phi$ (the scalar field whose fixed-point value generates the Planck scale) in addition to, of course,  the Starobinsky scalar $\zeta$ due to the $R^2$ term. Interestingly, however, its power spectrum is rather close to the observational bounds for inflation, giving us constraints within the ($n_s-r$) range to be within $n_s= 0.9618, r=0.0044$ from the latest PLANCK-BICEP data -- $n_s=0.9706, r=0.0026$ range (see Fig. \ref{nsrdata}). Furthermore, this also leads to the possibility to test this model with future observations with CMB from CMB-S4, Simon's Observatory, LiteBIRD and CMB-Bharat experiments.

Subsequently, we studied magnetogenesis in the context of scale-invariant inflation, described by the model \eqref{eq:axionRatra}. There, the conformal invariance of Maxwell's theory is broken by a coupling to the field $\zeta$ driving inflation, thus preventing vector perturbations from decaying adiabatically. We found that the combination of sawtooth coupling and helical magnetic fields provides current amplitudes and coherence lengths $(B_0, L_0)$ that are sufficient not only to seed galactic fields via cosmic dynamo but also satisfy the lower bounds from blazar observations. We have mainly focused on the case of scale-invariant (i.e., $k$-independent) magnetic and electric spectra at large scales/small wave vectors during the first stage of inflation, becoming blue-tilted at smaller scales in the second stage. We have further checked that the gauge field does not back-react on the inflationary evolution, and its contribution to tensor perturbations can be safely neglected. As a final remark, by contrast to what has been found in previous works, this model of magnetogenesis holds even at large and intermediate energy scales of inflation; moreover, its features gain a natural physical motivation since they are related to the specific dynamics of inflation arising in scale-invariant quadratic gravity.

\section*{References}
\bibliography{References}

\end{document}